\documentclass[superscriptaddress,twocolumn,showpacs,prb]{revtex4-1}
\usepackage[utf8]{inputenc} 
\usepackage{amsmath}
\usepackage{braket}
\usepackage{graphicx}
\usepackage{amsfonts}
\usepackage{pgfplots}
\usepackage{csquotes}
\usepackage{hhline}
\usepackage{amssymb}
\usepackage{listings}
\usepackage{color}

\definecolor{codegreen}{rgb}{0,0.6,0}
\definecolor{codegray}{rgb}{0.5,0.5,0.5}
\definecolor{codepurple}{rgb}{0.58,0,0.82}
\definecolor{backcolour}{rgb}{0.95,0.95,0.92}
 
\lstdefinestyle{mystyle}{
    backgroundcolor=\color{backcolour},   
    commentstyle=\color{codegreen},
    keywordstyle=\color{magenta},
    numberstyle=\tiny\color{codegray},
    stringstyle=\color{codepurple},
    basicstyle=\footnotesize,
    breakatwhitespace=false,         
    breaklines=true,                 
    captionpos=b,                    
    keepspaces=true,                 
    numbers=left,                    
    numbersep=5pt,                  
    showspaces=false,                
    showstringspaces=false,
    showtabs=false,                  
    tabsize=2
}
 
\usepackage{subfigure}

\usepackage{dcolumn}
\usepackage{tabularx}
\usepackage[colorlinks=true,linkcolor=blue,citecolor=blue,urlcolor=blue]{hyperref}
\usepackage{longtable}
\usepackage{braket}
\usepackage{float}
\newcolumntype{C}{>{\centering\arraybackslash}X}
\begin{document}
\title{Demonstrating Quantum Zeno Effect on IBM Quantum Experience}
\author{Subhashish Barik}
\email{subhashishbarik1995@gmail.com}
\affiliation{Department of Physical Sciences,\\ Indian Institute of Science Education and Research Kolkata, Mohanpur 741246, West Bengal, India}
\author{Dhiman Kumar Kalita}
\email{dhimankumarkalita@gmail.com}
\affiliation{Department of Physical Sciences,\\ Indian Institute of Science Education and Research Kolkata, Mohanpur 741246, West Bengal, India}

\author{Bikash K. Behera}
\email{bkb18rs025@iiserkol.ac.in}
\affiliation{Bikash's Quantum (OPC) Pvt. Ltd., Balindi, Mohanpur 741246, West Bengal, India}
\affiliation{Department of Physical Sciences,\\ Indian Institute of Science Education and Research Kolkata, Mohanpur 741246, West Bengal, India}
\author{Prasanta K. Panigrahi}
\email{pprasanta@iiserkol.ac.in}
\affiliation{Department of Physical Sciences,\\ Indian Institute of Science Education and Research Kolkata, Mohanpur 741246, West Bengal, India}

\begin{abstract}
\textbf{Abstract:}
Quantum Zeno Effect (QZE) has been one of the most interesting phenomena in quantum mechanics ever since its discovery in 1977 by Misra and Sudarshan [J. Math. Phys. \textbf{18}, 756 (1977)]. There have been many attempts for experimental realization of the same. Here, we present the first ever simulation of QZE on IBM quantum experience platform. We simulate a two-level system for Rabi-driven oscillation and then disturb the time evolution by intermediate repetitive measurements using quantum gates to increase the survival probability of the qubit in the initial state. The circuits are designed along with the added intermediate measurements and executed on IBM quantum simulator, and the outcomes are shown to be consistent with the predictions. The increasing survival probability with the number of intermediate measurements demonstrates QZE. Furthermore, some alternative explanations for the obtained results are provided which leads to some ambiguity in giving the exact reasoning for the observed outcomes.
\end{abstract}


\maketitle

\section{Introduction}

Quantum Zeno Effect (QZE) says that if we do repeated measurements on an unstable quantum system then we can slow down the quantum mechanical evolution of the system. This unusual effect after its discovery \cite{MSJMP1977} triggered many experimentalists to observe it practically. Many successful attempts \cite{IHBWPRA1990, NHCPRL1997, MDPRA2000, FGRPRL2001,NYKPRA2001, BHRWNTOC2002, HRBPKN2006, SMBCMKPPRL2006, SHCLCCSNC2014} have been reported so far in various experimental conditions. While the first attempt was to observe QZE in a two-level system with Rabi-driven oscillation \cite{IHBWPRA1990}, the later ones focussed on  multi-level systems \cite{SHCLCCSNC2014}, superconducting qubits \cite{MSKSPRB2006,KBMNSSNJP2015} etc. However, no attempts have been reported so far to simulate this effect on a quantum computer. Here, we address this issue by simulating QZE on IBM Quantum Experience (IBM QE).

We have tried to suppress the evolution of the initial state to a final state in a two-level Rabi-driven oscillation by frequent intermediate `measurements'. In other words, we have attempted to increase the probability of finding the qubit in the initial state when a unitary operator tends to evolve it to the final state by using frequent intermediate `measurements'. To be more precise, the attempt is to increase the survival probability (probability of surviving in the ground state) of the qubit during the two-level transition. It would be apt to mention that we have used the term `measurement' here in a loose sense and its actual interpretation with relevance to this article is discussed later in Section \ref{sec iii C}. At some places, we have also used the word `disturbance' for `measurement' as the former seems more generic. On the IBM QE, we have used $U3$ gates and CNOT gates as the intermediate disturbances while the the states $\ket{0}$ and $\ket{1}$ as the initial and final states of the desired two-level system respectively. From the knowledge of the Hamiltonian, H that would drive the two-level Rabi oscillation, we construct the time evolution operator, U and implement it using the $U3$ gate on IBM QE platform. To implement the frequent intermediate disturbances we change the parameters of the $U3$ gate and use them along with CNOT gates as shown in the Fig. \ref{qze_Fig3}.

\section{Theory\label{qnm_Sec2}}

The simplest non-trivial quantum mechanical example for explaining QZE would be the two-level system \cite{POSPD2014,GA2018}. Let us consider the well known Rabi oscillation for a two-level system caused by a generic Hamiltonian of the form $\hat{H}$ = $\Omega$($\ket{0}\bra{1}$ $+$ $\ket{1}\bra{0}$) where $\Omega$ is a time-independent constant. This Hamiltonian takes the qubit from state $\ket{0}$ to $\ket{1}$ and from state $\ket{1}$ to $\ket{0}$ and its matrix form looks like:
 
\begin{eqnarray}
\hat{H} ={\left( \begin{array}{cc} 0 & \Omega \\
\Omega & 0 \end{array} \right)}
\end{eqnarray}

Clearly then, $\hat{H} \ket{0}= \Omega \Ket{1}$ and $\hat{H} \ket{1}= \Omega \Ket{0}$. We can exponentiate this Hamiltonian to find the time evolution operator as:

\begin{eqnarray}
\hat{U}= e^{-i\hat{H}t}={\left( \begin{array}{cc} cos(\Omega t) & -isin(\Omega t) \\
-isin(\Omega t) & cos(\Omega t) \end{array} \right)}
\label{qze_Eq2}
\end{eqnarray}

It can be easily seen that if $\hat{U}$ acts on the initial state $\ket{0}$ then it would result in a superposition of states $\ket{0}$ and $\ket{1}$.
 
\begin{eqnarray}
\hat{U} \ket{0}=cos(\Omega t) \ket{0} -i sin(\Omega t) \ket{1}
\label{qze_Eq3}
\end{eqnarray}
 
Eq. \eqref{qze_Eq3} clearly shows that the survival probability, $P_s$ of staying in state $\ket{0}$ after some time `\textit{t}' is $cos^2(\Omega t)$. We can further write:

\begin{eqnarray}
P_s =cos^2(\Omega t) =\frac{1}{2} (1+cos(2 \Omega t))\nonumber\\
=\frac{1}{2} (1+1-\frac{(2 \Omega t)^2}{2!} + \frac{(2 \Omega t)^4}{4!} - \ldots )
\end{eqnarray}
 
Neglecting the higher order terms by considering $t$ to be small we obtain:

\begin{eqnarray}
P_s = \frac{1}{2}\bigg (2-\frac{(2 \Omega t)^2}{2}\bigg)=1-\Omega^2 t^2 
\end{eqnarray}

Now, if we divide $t$ to `$\textit{n}$' intervals and `measure' after each interval then after the final $n^{th}$ interval's `measurement' the survival probability becomes:

\begin{eqnarray}
P_s= \bigg(1- \Omega^2 \bigg(\frac{t}{n}\bigg)^2\bigg)^n
\label{qze_Eq6}
\end{eqnarray}

Considering $\frac{t}{n}$ to be very small as compared to 1 we can further write Eq. \eqref{qze_Eq6} as:

\begin{eqnarray}
P_s=1-\frac{\Omega ^2 t^2}{n}
\label{qze_Eq7}
\end{eqnarray}

Eq. \eqref{qze_Eq7} clearly shows that the survival probability increases with the number of intervals or the number of intermediate `measurements' $n$ and with increasing value of $n$ the survival probability tends towards unity.

\begin{equation}
P_s \propto n \ and \ \lim_{n \to \infty} P_s = 1
\end{equation}

\section{Implementation on IBM Quantum Experience \label{qnm_3}}

\subsection{Setting up the basic circuit}
 
When we think of implementing the above theoretical formulation on IBM QE the primary task is to prepare the desired initial state and the unitary time evolution operator. IBM QE initializes all the qubits in the state $\ket{0}$, hence the first task is done. For the second task, we use the $U3$ gate provided on IBM QE; we set the parameters $\theta$, $\phi$ and $\lambda$ as per our requirement to simulate $\hat{U}$ of Eq. \eqref{qze_Eq2}. The $U3$ gate on IBM QE has the following form:

\begin{eqnarray}
U3(\theta,\phi,\lambda)={\left( \begin{array}{cc} cos(\theta/2) & -e^{i\lambda}sin(\theta/2) \\
e^{i\phi}sin(\theta/2) & e^{i(\lambda+\phi)}cos(\theta/2) \end{array} \right)}
\end{eqnarray}

\begin{figure}[h]
   \centering
   \includegraphics[scale=0.6]{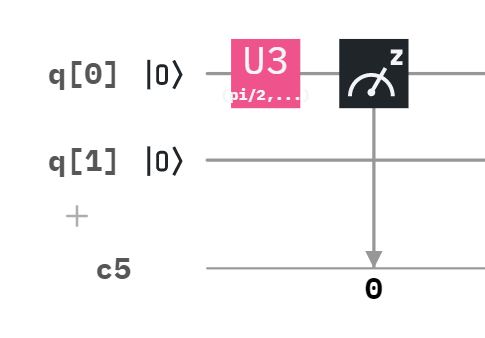}
   \caption{Circuit describing the operation of $U3$ on $q[0]$} qubit where, the $U3$ has the value of $\theta=\pi/2, \phi=-\pi/2$ and $\lambda=\pi/2$. The measurement gate measures only the first qubit in the computational \{$\ket0$, $\ket1$\} basis.
   \label{qze_Fig1}
\end{figure}

Thus, by choosing the parameters $\phi=-\pi/2, \lambda=\pi/2$ and $\theta=2\Omega t$, we make $U3$ equal to $\hat{U}$. Now, we are ready to operate $\hat{U}$ on state $\ket{0}$ i.e., to implement $U3$ on qubit $q[0]$ as shown in Fig. \ref{qze_Fig1}.

\begin{figure}[h]
   \centering
   \includegraphics[scale=0.2]{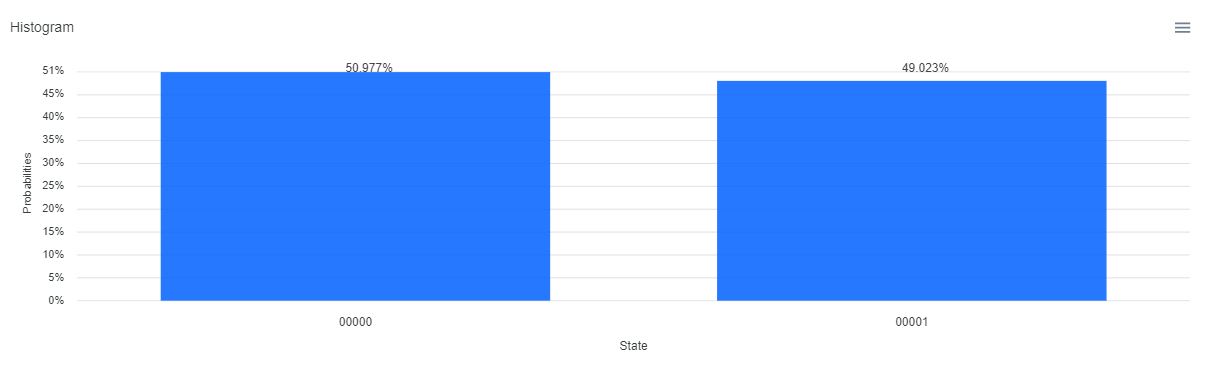}
   \caption{Measurement outcomes of $q[0]$ after the operation of $U3$ on ibmq qasm simulator for 8192 shots, which gives precisely 50.464\% probability for state $\ket0$ and 49.536\% for state $\ket1$.}
   \label{qze_Fig2}
\end{figure}

The measurement gate added to the circuit after $U3$ gate in Fig. \ref{qze_Fig1} measures the qubit $q[0]$. If we choose $\theta=\pi/2$ then $q[0]$ is measured to be found in $\ket{0}$ and $\ket{1}$ with roughly equal probability of 50\% as shown in the Fig. \ref{qze_Fig2} which is in fact what we would expect theoretically from Eq. \eqref{qze_Eq3}.

\subsection{Adding the intermediate measurements}
\label{seciiib}

\begin{figure}[h]
   \centering
   \includegraphics[scale=0.5]{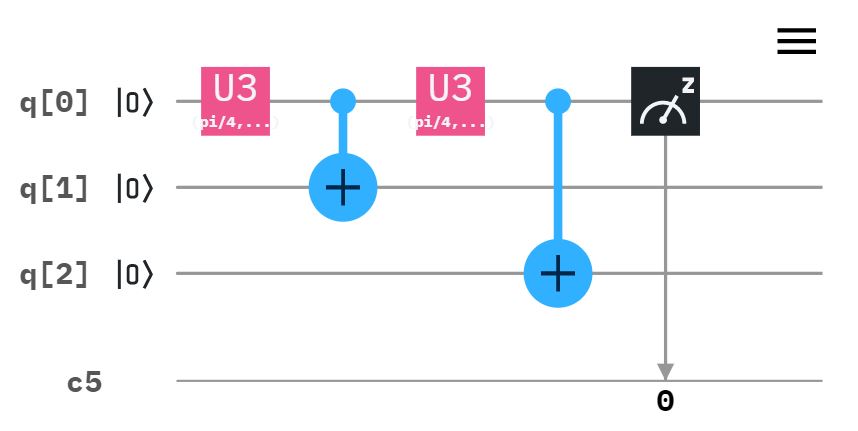}
   \caption{Quantum circuit for \textit{n}=2. The circuit describes the operation of two \textit{U3} gates and two \textit{CNOT} gates on $q[0]$ where, each of the $U3$ has the value of $\theta=\pi/4, \phi=-\pi/2$ and $\lambda=\pi/2$. The measurement gate measures only the first qubit in the computational ($\ket0$, $\ket1$) basis.}
   \label{qze_Fig3}
\end{figure}

\begin{figure}[h]
   \centering
   \includegraphics[scale=0.2]{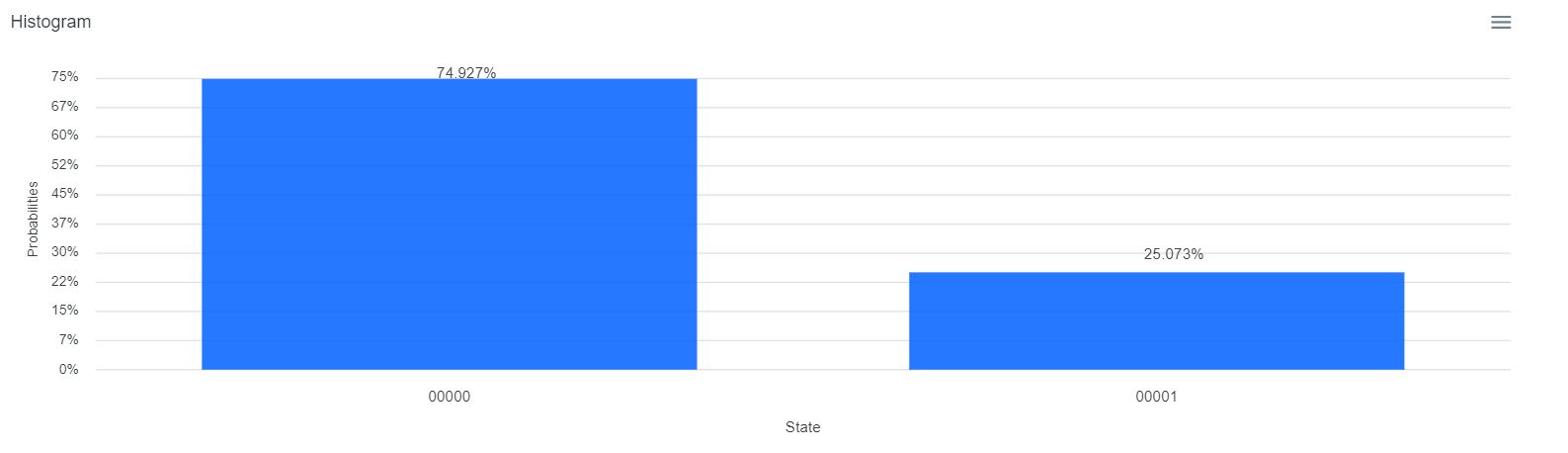}
   \caption{Measurement outcomes of $q[0]$ after the operation of two $U3$ gates with parameters $\theta=\pi/4$, $\phi=-\pi/2$ and $\lambda=\pi/2$ and two $CNOT$ gates on ibmq qasm simulator for 8192 shots, which gives precisely 74.927\% probability for state $\ket0$ and 25.073\% for state $\ket1$.}
   \label{qze_Fig4}
\end{figure}

In the next step, we proceed towards observing the QZE. Hence, we add the intermediate disturbances dividing the `measurement' interval to equal halves. For $n$=2, we use two $U3$ gates each having $\theta$=$\pi$/4. The values of $\phi$ and $\lambda$ are kept unchanged. Further, we add one $CNOT$ gate right after each $U3$ gate and at the end we put the measurement box in the qubit line of $q[0]$. For making the above circuit, we use the fact that the use of two $U3$ gates having $\theta=\pi/4$ is equivalent to the use of a single $U3$ gate with $\theta=\pi/2$; this concept is further explained in the Appendix \ref{Appendix A}. The resulting circuit is described in Fig. \ref{qze_Fig3} and the outcomes of the measurement of this circuit are shown in Fig. \ref{qze_Fig4}.

We have used here the result that operating one $U3$ gate with $\theta$=$\pi/2$ is equivalent to operating two $U3$ gates with $\theta$=$\pi/4$ and to generalize it we can say that operating one $U3$ gate with $\theta$=$\pi/m$ is equal to operating $n$ $U3$ gates sequentially with  $\theta$=$(\pi/m)/n$=$\pi/mn$.

\begin{figure}[h]
   \centering
   \includegraphics[scale=0.4]{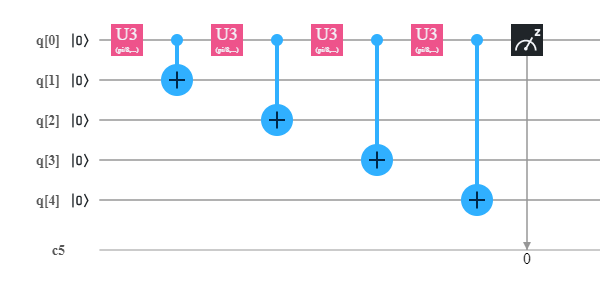}
   \caption{Quantum circuit for $n$=2. The circuit describes the operation of four $U3$ gates and four $CNOT$ gates on $q[0]$ where, each of the $U3$ has the value of $\theta=\pi/8, \phi=-\pi/2$ and $\lambda=\pi/2$. The measurement gate measures only the first qubit in the computational ($\ket0$, $\ket1$) basis}
   \label{qze_Fig5}
\end{figure}

We can clearly see the effect of adding one intermediate measurement; the survival probability goes to roughly 75\% from 50\%. Next, we keep increasing the value of $n$ in our quantum circuit gradually and with each increment of $n$ we put one extra $U3$ gate along with one extra $CNOT$ gate by taking $\theta$=$\pi/2n$. We put the measurement gate at the end. For example, for $n$=4, we take four $U3$ gates each having $\theta$=$\pi/8$ and followed by a CNOT as shown in Fig. \ref{qze_Fig5}.

\begin{figure}[h]
   \centering
   \includegraphics[scale=0.4]{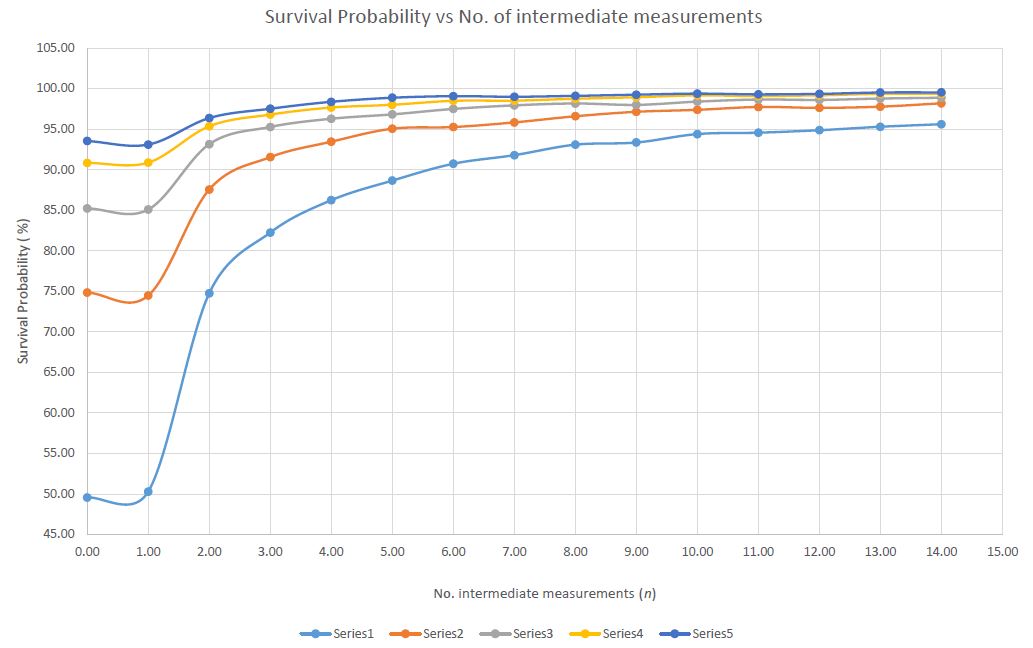}
   \caption{The plot of survival probability vs number of intermediate measurements $n$. The series 1, 2, 3, 4 and 5 correspond to the $\theta$ values $\pi/2$, $\pi/3$, $\pi/4$, $\pi/5$ and $\pi/6$ respectively. The curve for a higher $\theta$ value saturates faster as its initial survival probability (for $n$=1) of it is higher.}
   \label{qze_Fig6}
\end{figure}

We prepare different quantum circuits for each $n$ and take measurements for $n$=2 to 14. We then plot the obtained survival probabilities against their corresponding $n$. We do the above process for five different sets of $\theta$ values; $\theta$= $\pi/2$, $\pi/3$, $\pi/4$, $\pi/5$, $\pi/6$ and then plot the curves of survival probability for each set of $\theta$ value. The resulting plot is given in Fig. \ref{qze_Fig6}.

It is very conspicuous from Fig. \ref{qze_Fig6} that the survival probability increases continuously with $n$ and then saturates close to 100\% for higher values of $n$. This demonstrates the QZE.

\subsection{The meaning of `measurement' here}
\label{sec iii C}

The question of what defines a `measurement' in the context of QZE has been debated vigorously in the literature \cite{IHBWPRA1990, ICS2019, HWAP1997, SWAJP1982, KSPR2005, MA2005, SRMP2005}. Different researchers have used different methods to show measurements in the context of QZE \cite{HDWCPRA2019, BPA2019}.  Sometimes using the term `intermediate disturbance' seems more convenient while sometimes the term `projective measurement' makes more sense. To avoid going into the controversial interpretation of its actual meaning we would like to call it \textbf{deferred and implicit measurement} \cite{C2019}. The principle of deferred and implicit measurement says that if we leave some quantum wires untouched we can assume they are measured \cite{NCQCQI2010,CTQC2012}. Based on this principle, we can treat each intermediate $CNOT$ gate as a valid intermediate measurement and thus it is these intermediate $CNOT$ gates which causes the increase in the survival probability.

\section{Conclusion \label{qnm_Sec6}}
We have shown here that the survival probability of staying in state $\ket{0}$ increases with the number of intermediate measurements; more particularly `deferred and implicit measurements'. In other words, we have suppressed the transition of the qubit from state $\ket{0}$ to $\ket{1}$. From this point of view, the observed behaviour in Fig. \ref{qze_Fig6} seems to demonstrate QZE. However, looking at the theoretical details of the quantum circuits as done in Appendix \ref{Appendix B}, the observed behaviour just looks like the outcomes of the trivial calculations for the operation of $U3$ and $CNOT$ gates. This creates an ambiguity in whether we can call it as a valid demonstration of QZE or not. Thus, we would like to conclude here with this open question and suggest a deeper look into the matter in future work. Moreover, we propose extending this methodology of adding `deferred and implicit measurements' to higher level systems. That would require us to come up with clever ways to simulate the time evolution operator using the gates of the IBM QE library and the multi-energy levels using multiple-qubit states. One particular implementation could be of a three level system's dynamics where we would like to confine the transitions to only the lower energy levels by suppressing the transition to the highest energy level. In fact, this could be the simulation of a possible resolution to the leakage problem in superconducting quantum computing architectures \cite{GA2018}.

\section*{Acknowledgments}
\label{qlock_acknowledgments}
S.~B. and D.~K.~K. would like to thank Bikash's Quantum (OPC) Pvt. Ltd. and IISER Kolkata respectively for providing hospitality during the course of the project work. S.~B. would also like to thank Prof. I.~S. Tyagi of Indian Institute of Technology Ropar for motivating and giving him the opportunity to explore the Quantum Zeno Effect during his Master thesis. B.~K.~B. acknowledges the support of IISER-K Institute fellowship. The authors acknowledge the support of IBM Quantum Experience. The views expressed are those of the authors and do not reflect the official policy or position of IBM or the IBM Quantum Experience team.

\section*{Competing interests}
The authors declare no competing financial as well as non-financial interests.

\appendix

\section{The reasoning using which we decompose a single $U3$ to multiple $U3$.}
\label{Appendix A}

In Section \ref{seciiib}, we have used the result that operating one $U3$ gate with $\theta=\pi/2$ is equivalent to operating two $U3$ gates with $\theta=\pi/4$ and to generalize it we can say that operating one $U3$ gate with $\theta=\pi/m$ is equal to operating $n$ $U3$ gates sequentially with $\theta=(\pi/m)/n=\pi/mn$.

\begin{figure}[h]
   \centering
   \includegraphics[scale=0.3]{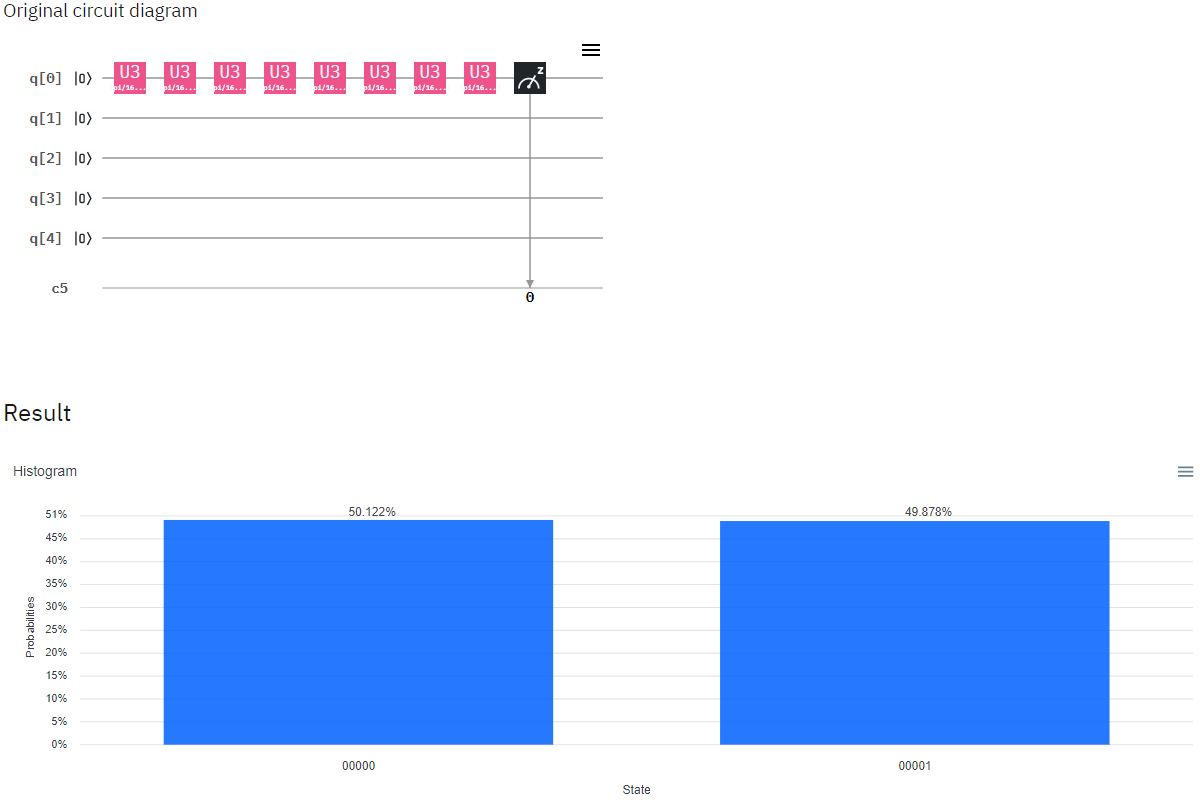}
   \caption{Plot showing the outcomes of measurement for eight \textit{U3} gates each having $\theta=\pi/16$. After the comparision with Fig. \ref{qze_Fig2}, the 50\% probability of getting state $\Ket{0}$ here shows that using eight \textit{U3} gates of $\theta=\pi/16$ is equivalent to using one \textit{U3} gate of $\theta=\pi/2$.}
   \label{qze_Fig7}
\end{figure}

To give specific examples, we start with $m$=2 and $n$=8. This particularly refers to using 8 \textit{U3} gates sequentially each having $\theta=\pi/2$. The circuit and its measurement outcomes are shown in Fig. \ref{qze_Fig7}. It can be seen that the survival probability is found to be 50.122\% which is in close agreement with the result obtained in Fig. \ref{qze_Fig7} i.e., for a single \textit{U3} gate with $\theta=\pi/2$.

\begin{figure}[h]
   \centering
   \includegraphics[scale=0.3]{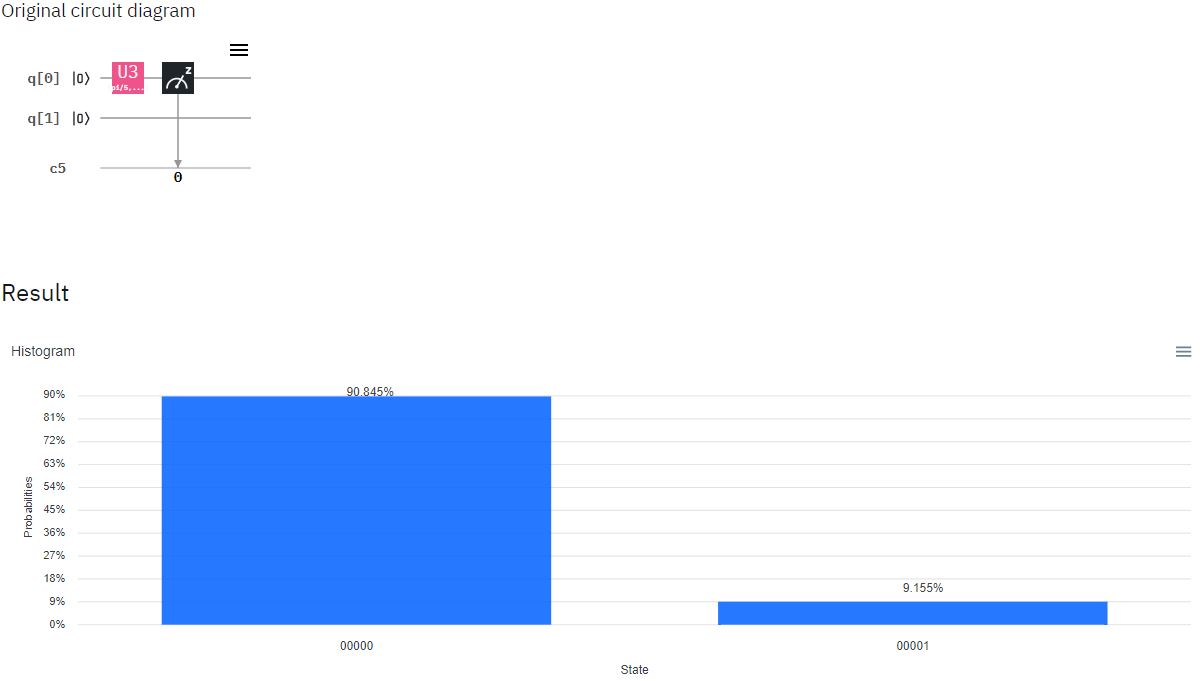}
   \caption{Plot showing the outcomes of measurement for one \textit{U3} gate having $\theta=\pi/5$. It shows the probability for obtaining state $\ket{0}$ to be around 90\% }
   \label{qze_Fig8}
\end{figure}

\begin{figure}[h]
   \centering
   \includegraphics[scale=0.3]{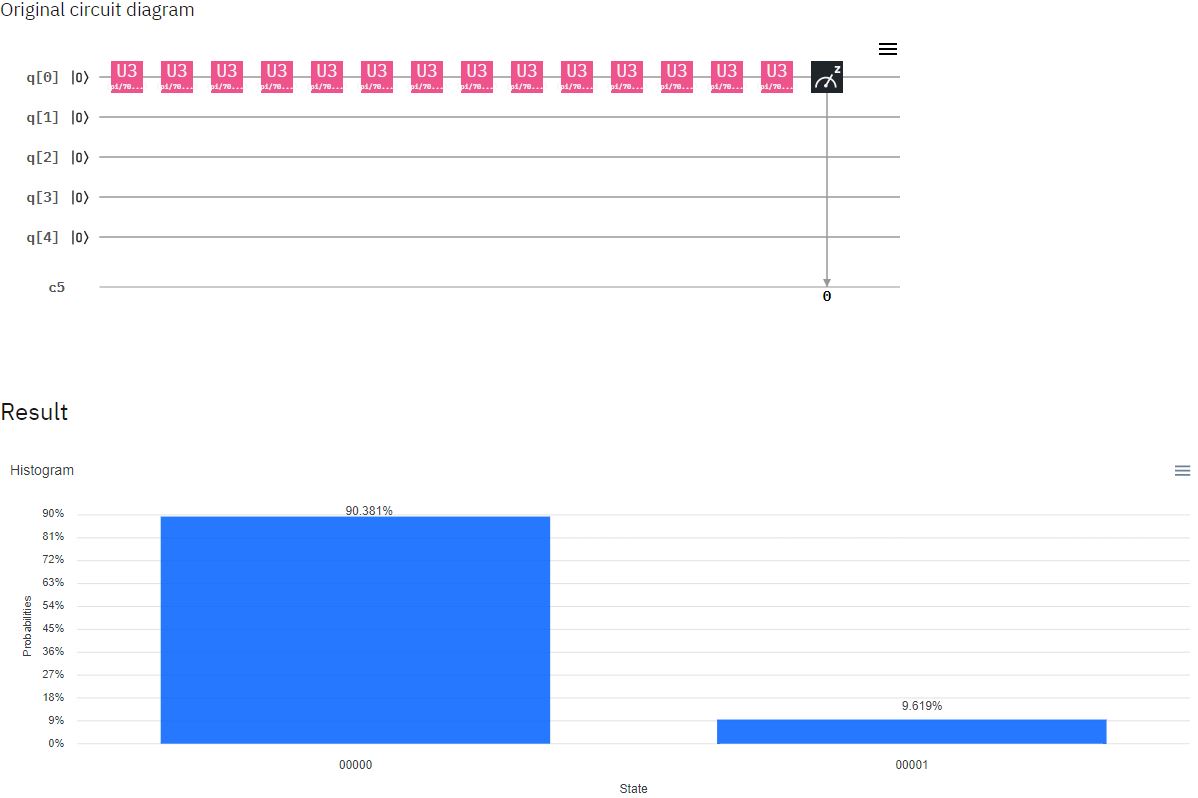}
   \caption{Plot showing the outcomes of measurement for fourteen \textit{U3} gates each having $\theta=\pi/70$. After comparision with Fig. \ref{qze_Fig8}, the 90\% probability of getting state $\Ket{0}$ here shows that using fourteen \textit{U3} gates of $\theta=\pi/70$ is equivalent to using one \textit{U3} gate of $\theta=\pi/5$}
   \label{qze_Fig9}
\end{figure}

Next, we take the example for $m$=5 and $n$=14. This particularly refers to using 14 \textit{U3} gates sequentially each having $\theta=\pi/5$. The circuit and its measurement outcomes are shown in Fig. \ref{qze_Fig9}. It can be seen that the survival probability is found to be 90.381\% which is in close agreement with the result obtained in Fig. \ref{qze_Fig8} i.e., for a single \textit{U3} gate with $\theta=\pi/5$ which gives the survival probability as 90.485\%. In this way we can verify this for all different values of $m$ and $n$ on IBM QE.

\section{The theory behind the operation of intermediate $U3$ and $CNOT$ gates}
\label{Appendix B}

From Eq. \eqref{qze_Eq3} we can find the form of $U3$ matrix for $\phi=-\pi/2$ and $\lambda=\pi/2$ to be:

\begin{eqnarray}
U3(\theta,-\pi/2,\pi/2)={\left( \begin{array}{cc} cos(\theta/2) & -isin(\theta/2) \\
-isin(\theta/2) & cos(\theta/2) \end{array} \right)}
\label{qze_EqB1}
\end{eqnarray}

From Eq. \eqref{qze_EqB1} we can infer that:
\begin{eqnarray}
U3\ket{0}= cos(\theta/2)\ket{0}-isin(\theta/2)\ket{1}\\
U3\ket{1}= cos(\theta/2)\ket{1}-isin(\theta/2)\ket{0}
\end{eqnarray}

For some simplicity let us write,
\begin{eqnarray}
U3\ket{0}= \alpha\ket{0}+\beta\ket{1}\\
U3\ket{1}= \beta\ket{0}+\alpha\ket{1}
\end{eqnarray}

That is to say, we have:
\begin{eqnarray}
\alpha=\cos(\theta/2)\\
\beta=-i\sin(\theta/2)
\end{eqnarray}

with $|\alpha|^2+|\beta|^2=1$. Using this formalism, for Fig. \ref{qze_Fig3} i.e., for quantum circuit with two $U3$ gates, we can analytically obtain the result for the survival probability of obtaining state $\ket{0}$ as explained below:

\textbf{Step 1}: $U3$ operates on $\ket{0}$ 
\begin{eqnarray}
\ket{\psi_1}=U3\Ket{0}=\alpha\ket{0}+\beta\ket{1}
\end{eqnarray}

\textbf{Step 2}: $CNOT$ acts on $\ket{\psi_1}\otimes\ket{0}$

\begin{eqnarray}
\ket{\psi_2}&=&CNOT(\alpha\ket{0}+\beta\ket{1})\otimes\Ket{0}\nonumber\\
&=&CNOT(\alpha\ket{00}+\beta\ket{10})\nonumber\\
&=&\alpha\ket{00}+\beta\ket{11}
\end{eqnarray}

\textbf{Step 3}: $U3$ operates on $\ket{\psi_2}$
\begin{eqnarray}
\ket{\psi_3}&=&U3(\alpha\ket{00}+\beta\ket{11})\nonumber\\
&=&\alpha(U3\ket{0})\otimes\ket{0}+\beta(U3\ket{1})\otimes\ket{1}\nonumber\\
&=&\alpha(\alpha\ket{0}+\beta\ket{1})\otimes\ket{0}+\beta(\alpha\ket{1}+\beta\ket{0})\otimes\ket{1}\nonumber\\
&=&\alpha^2\ket{00}+\alpha\beta\ket{10}+\alpha\beta\ket{11}+\beta^2\ket{01}
\end{eqnarray}

\textbf{Step 4}: $CNOT$ acts on $\ket{\psi_3}\otimes\ket{0}$
\begin{eqnarray}
\Ket{\psi_4}&=&CNOT(\ket{\psi_3}\otimes\ket{0})\nonumber
\\
\ket{\psi_4}&=&CNOT(\alpha^2\ket{00}+\alpha\beta\ket{10}+\alpha\beta\ket{11}\nonumber\\
&&+ \beta^2\ket{01})\otimes\ket{0}\nonumber\\
&=&CNOT(\alpha^2\ket{000}+\alpha\beta\ket{100}+\alpha\beta\ket{110}\nonumber\\&&
+\beta^2\ket{010})\nonumber\\
&=&\alpha^2\ket{000}+\alpha\beta\ket{101}+\alpha\beta\ket{111}\nonumber\\
&&+\beta^2\ket{010}
\end{eqnarray}

\textbf{Step 5}: Measurement on first qubit of $\ket{\psi_4}$

This gives the probability of getting $\ket{0}$ for the first qubit as:

\begin{eqnarray}
P_s=|\alpha^2|^2+|\beta^2|^2= \alpha^4+\beta^4
\end{eqnarray}

For Fig. \ref{qze_Fig3}, we have $\theta=\pi/2$ this result comes out to be:
$P_s=cos^4(\theta/2)+sin^4(\theta/2)=cos^4(\pi/4)+sin^4(\pi/4)$=0.7500 theoretically and 0.7475 from IBM QE (as plotted in Fig. \ref{qze_Fig6}). A similar treatment for three $U3$ gates of $\theta=\pi/6$ yields the probability of obtaining state $\ket{0}$ to be $|\alpha|^6 +3|\alpha|^2|\beta|^4=cos^6(\pi/12)+3cos^2(\pi/12))sin^4(\pi/12)$. This result comes out to be 0.8247 theoretically and 0.8223 from IBM QE (as plotted in Fig. \ref{qze_Fig6}). So, we can see a close agreement between the theoretically predicted outcomes and the ones given by IBM QE. We can extend this argument to all other values of $\theta$ and $n$ as well.


\begin{thebibliography}{10}
\expandafter\ifx\csname url\endcsname\relax
\def\url#1{\texttt{#1}}\fi
\expandafter\ifx\csname urlprefix\endcsname\relax\def\urlprefix{URL }\fi
\providecommand{\bibinfo}[2]{#2}
\providecommand{\eprint}[2][]{\url{#2}}

\section*{References}

\bibitem{MSJMP1977}B. Misra and E.~C.~G. Sudarshan, The Zeno's paradox in quantum theory, J. Math. Phys. \textbf{18}, 756 (1977).

\bibitem{IHBWPRA1990}W.~M. Itano, D.~J. Heinzen, J.~J. Bollinger, and D.~J. Wineland, Quantum Zeno Effect, Phys. Rev. A \textbf{41}, 2295 (1990).

\bibitem{NHCPRL1997}B. Nagels, L.~J.~F. Hermans, and P.~L. Chapovsky, Quantum Zeno Effect Induced by Collisions, Phys. Rev. Lett. \textbf{79}, 3097 (1997).

\bibitem{MDPRA2000}K. Molhave, and M. Drewsen, Formation of translationally cold MgH+ and MgD+ molecules in an ion trap, Phys. Rev. A \textbf{62}, 011401(R) (2000).

\bibitem{FGRPRL2001} M.~C. Fischer, B. Gutierrez-Medina, and M.~G. Raizen, Observation of the quantum Zeno and anti-Zeno effects in an unstable system, Phys. Rev. Lett. \textbf{87}, 040402 (2001).

\bibitem{NYKPRA2001}T. Nakanishi, K. Yamane, and M. Kitano, Absorption-free optical control of spin systems: The quantum Zeno effect in optical pumping, Phys. Rev. A \textbf{65}, 013404 (2001).

\bibitem{BHRWNTOC2002}C. Balzer, T.Hannemann, D. ReiB, C. Wunderlich, W. Neuhauser, and P.~E. Toschek, A relaxationless demonstration of the Quantum Zeno paradox on an individual atom, Opt. Commun. \textbf{211}, 235 (2002).

\bibitem{HRBPKN2006}O. Hosten, M.~T. Rakher, J.~T. Barreiro, N.~A. Peters, and P.~G. Kwiat, Counterfactual quantum computation through quantum interrogation, Nature \textbf{439}, 949 (2006).
 
\bibitem{SMBCMKPPRL2006}E.~W. Streed, J. Mun, M. Boyd, G.~K. Campbell, P. Medley, W. Ketterle, and D.~E. Pritchard, Large atom number Bose-Einstein condensate machines, Phys. Rev. Lett. \textbf{97}, 260402 (2006).

\bibitem{SHCLCCSNC2014}F. Schafer, I. Herrera, S. Cherukattil, C. Lovecchio, F.~S. Cataliotti, F. Caruso, and A. Smerzi, Experimental realization of quantum zeno dynamics, Nat. Commun. \textbf{5}, 3194 (2014).

\bibitem{MSKSPRB2006}Y. Matsuzaki, S. Saito, K. Kakuyanagi, and K. Semba, Quantum Zeno effect with a superconducting qubit, Phys. Rev. B \textbf{82}, 180518(R) (2010).

\bibitem{KBMNSSNJP2015}K. Kakuyanagi, T. Baba, Y. Matsuzaki, H. Nakano, S. Saito, and K. Semba, Observation of quantum Zeno effect in a superconducting flux qubit, New J. Phys. \textbf{17}, 063035 (2015).

\bibitem{POSPD2014}S. Pascazio, All You Ever Wanted to Know About the Quantum Zeno Effect in 70 Minutes, Open Sys. Inf. Dyn. \textbf{21}, 1440007 (2014).

\bibitem{GA2018}A.~A. Galiautdinov, Quantum Zeno effect: A possible resolution to the leakage problem in superconducting quantum computing architectures, arXiv:1805.06877, (2018).

\bibitem{ICS2019}W.~M. Itano, The quantum Zeno paradox, 42 years on*, SPECIAL SECTION: E.~C.~G. SUDARSHAN, Current Science \textbf{116}, 2 (2019).

\bibitem{HWAP1997}D. Home, M.~A.~B. Whitaker, A Conceptual Analysis of Quantum Zeno; Paradox, Measurement, and Experiment, Annal. Phys. \textbf{258}, 237 (1997).

\bibitem{SWAJP1982}I. Singh and M.~A.~B. Whitaker, Role of the observer in quantum mechanics and the Zeno paradox, Am. J. Phys. \textbf{50}, 882 (1982).

\bibitem{KSPR2005}K. Koshino, and A. Shimizu, Quantum Zeno effect by general measurements, Phys. Rep. \textbf{412}, 191 (2005).

\bibitem{MA2005}A.~N. Mitra, Foundations Of Quantum Theory Revisited, arXiv:quant-ph/0510223, (2005).

\bibitem{SRMP2005}M. Schlosshauer, Decoherence, the measurement problem, and interpretations of quantum mechanics, Rev. Mod. Phys. \textbf{76}, 1267 (2005).

\bibitem{HDWCPRA2019}S. He, L.W. Duan, C. Wang, Q.H. Chen, Quantum Zeno Effect in a circuit-QED system, Phys. Rev. A, \textbf{99}, 052101, (2019)

\bibitem{BPA2019}S. Belan, V. Parfenyev, Optimal Measurement Protocols in Quantum Zeno Effect, arXiv:1909.03226 [cond-mat.stat-mech], 
(2019)

\bibitem{C2019}C. Calcluth, \url{https://phys.cam/2019/07/quantum-zeno-effect/}, (2019).

\bibitem{NCQCQI2010}M. Nielsen, and I.~L. Chuang, ``4.4 Measurement". Quantum Computation and Quantum Information: 10th Anniversary Edition, Cambridge University Press, 186, (2010).

\bibitem{CTQC2012}O.~A. Cross, ``5.2.2 Deferred Measurement" Topics in Quantum Computing, 348, (2012).

\end{thebibliography}
\end{document}